# VACUUM WAVES

Paul S. Wesson

Department of Physics and Astronomy, University of Waterloo, Waterloo, Ontario N2L 3G1, Canada

Herzberg Institute of Astrophysics, National Research Council, Victoria, B.C. V9E 2E7, Canada

Abstract:  As an example of the unification of gravitation and particle physics, an exact solution of the five-dimensional field equations is studied which describes waves in the classical Einstein vacuum.  While the solution is essentially 5D in nature, the waves exist in ordinary 3D space, and may provide a way to test for an extra dimension.

# VACUUM WAVES

## 1.  Introduction

Five-dimensional relativity is the simplest extension of general relativity that might be expected to show the low-energy effects of even higher dimensions connected with the physics of elementary particles.  Below, I will examine the properties of an exact solution of the 5D field equations which is of special interest.  It describes the classical Einstein vacuum, with periodic perturbations which exist in ordinary 3D space (Section 2).  These waves have unusual properties, so it is difficult to classify them.  They require the presence of a vacuum and have phase velocities which can exceed the speed of light, implying that they are not gravitational waves but more resemble de Broglie waves.  If these vacuum waves were present in the early universe or persist today, they may provide a way to test for an extra dimension.

## 2.  An Exact Solution with Vacuum Waves

The field equations of five-dimensional relativity are commonly taken to be given by the Ricci tensor as $R_{AB} = 0$ ($A, B = 0,1,2,3,4$ for time, ordinary space and the extra coordinate). These 5D equations actually contain Einstein's 4D equations, by Campbell's embedding theorem [1]. It is a corollary of this theorem that when the 5D metric takes the so-called canonical form, it embeds all solutions of general relativity which are empty of ordinary matter but contain finite vacuum energy as measured by the cosmological constant [2].  The latter parameter $\Lambda$ is positive or negative, depending on whether the extra dimension is spacelike or timelike.  For both cases, the equation of state of the vacuum in terms of its effective energy density and pressure is $p_v = -\rho_v = -\Lambda / 8\pi$. Here units have been chosen which render the speed of light $c$ and the gravitational constant $G$ unity, though these quantities and Planck's constant $h$ will be made explicit later in order to aid physical understanding. The extra coordinate will be labelled $x^4 = l$, to avoid confusion with the usual coordinates of spacetime $x^\alpha = t, xyz$. Other notation is standard.

There are many solutions known of the 5D field equations $R_{AB} = 0$, including several in canonical form [3-5]. The 5D Schwarzschild-de Sitter solution is one such, which ensures the agreement of 5D relativity with the classical tests of Einstein's theory. But for present purposes, consider the following solution:

$$dS^2 = \frac{l^2}{L^2}\left\{dt^2 - \exp\left[\pm\frac{2i}{L}\left(t + \alpha x\right)\right]dx^2 - \exp\left[\pm\frac{2i}{L}\left(t + \beta y\right)\right]dy^2\right.$$

$$\left. - \exp\left[\pm\frac{2i}{L}\left(t + \gamma z\right)\right]dz^2\right\} + dl^2 \quad . \qquad (1)$$

This describes a wave propagating through ordinary 3D space, where the frequency $f = 1/L$ is fixed by the solution. The wave-numbers $k_x = \alpha / L, \quad k_y = \beta / L, \quad k_z = \gamma / L$ along the $x$, $y$, $z$ axes are written in terms of the dimensionless constants $\alpha$, $\beta$, $\gamma$ which are arbitrary. The phase velocity of the wave along (say) the x-axis is $c / \alpha$, and is also arbitrary. By evaluating the Einstein tensor from the 4D part of (1), it may be shown by some algebra that the equations of general relativity are satisfied, with a cosmological constant $\Lambda = -3 / L^2$. Accordingly, (1) describes waves travelling in a classical vacuum with positive pressure.

Other properties of (1) may be revealed by using one of the software packages currently available. In fact, while (1) was found originally by solving the field equations by hand, the quickest way to verify it is by computer.

Perhaps the most striking feature of (1) is that it allows phase velocities that are, in a formal sense, greater than the speed of light. This because, as noted above, the phase speed is $c/\alpha$ and $\alpha$ can be arbitrarily small. It should be recalled that this does not necessarily conflict with causality, as long as the waves concerned do not carry conventional information [6]; and there is a (somewhat unusual) interpretation of the Lorentz transformations due to Rindler that allows such speeds [7]. It should also be recalled that the solution (1) requires $\Lambda < 0$ with vacuum density $\rho_v < 0$, and such a medium is unlike any form of ordinary matter. All waves in ordinary matter have velocities given by an expression of the sort $w = \sqrt{K/\rho}$, where $K$ is a constant that depends on microscopic physics (the bulk modulus for fluids and the shear modulus for solids). This is discussed in standard texts, a few of which point out that a value $K < 0$ would imply that an increase in the ambient pressure causes an increase in the volume of an element of the material, or equivalently a decrease in the density. This behaviour is the opposite of that observed in ordinary matter. However, it is exactly the behaviour consistent with the equation of state $p_v = -\rho_v c^2$ of the vacuum when the pressure is positive and the density is negative. In understanding the possibility of superluminal velocities as given by the solution (1) of 5D relativity, it is also useful to consider 4D wave mechanics. There, a wave with phase speed $w$ has associated with it a particle with ordinary speed $v$, and the application of Planck's law implies that the two are related by $vw = c^2$. Then if $v < c$ it follows necessarily that $w > c$.

In fact, the relation $vw = c^2$ is named after de Broglie and is well known in quantum theory. Accordingly, it is logical to inquire into the connection between (1) and de Broglie waves.

### 3. Vacuum Waves and De Broglie Waves

In the preceding section, an exact 5D solution (1) was given which is characterized by waves which in ordinary 3D space resembles de Broglie waves. The question arises of whether this correspondence can be made exact, and if so what other implications arise for wave mechanics.

The theoretical basis of de Broglie waves was stated by him in 1923. He postulated that any massive particle has associated with it a wave, in analogy with a photon's electromagnetic wave, in a way that respects the Lorentz transformations and Planck's law. The direction of motion of the particle is the same as the normal to the wave-front. Let the mass and velocity of the particle be $m$, $v$ and the frequency and velocity of the wave be $f$, $w$. Then in terms of the 4-vectors for the particle and the wave, there is a match of the magnitude of the quantities concerned, via de Broglie's equation: $m(v,1) = hf(1/w, 1/c^2)$. Equating components in de Broglie's equation gives $mv = hf/w$ and $m = hf/c^2$. The second of these relations is simply a statement of the equality of the energy of the particle and the energy of the wave: $E = mc^2 = hf$. The first relation when divided by the second, gives $vw = c^2$.

To see how this relation compares with the wave of the solution (1), consider the element of proper distance in the latter along the x-axis, given by $d\overline{x} = \exp\left[i\left(ft + k_x x\right)\right]dx$. The frequency is $f = c/L$. By Planck's law, the energy of the wave is equivalent to the mass $m$ of the associated particle, $E = hf = hc / L = mc^2$, so $L = h / mc$. That is, the size of the 4D potential well in (1) equals the Compton wavelength of the particle. The momentum of the particle $p_x = m v_x$ is inversely proportional to the wavelength, so the wave-number $k_x$ is directly proportional to $p_x$, and can be written in the correct dimensional form as $k_x = \left( mc / h \right) \left( v_x / c \right) = v_x / cL$. Recalling that the frequency is $f = c / L$, the phase velocity of the wave along the x-axis is $w_x = f / k_x = \left( c / L \right) \left( cL / v_x \right) = c^2 / v_x$. Thus along each axis, the velocities of the particle and the wave are connected by de Broglie's relation $vw = c^2$.

In standard wave mechanics, the problematical nature of the medium which supports the waves is effectively avoided, by introducing a complex wave function $\psi$ that is abstract and makes no direct reference to conventional properties of matter. The theory employs operators, which act on the time and space parts of the 4D metric to produce the energy and momenta of a particle, whose mass is then given by a relation in $\psi$ called the Klein-Gordon equation. Since wave mechanics is used to successfully model the interactions of particles, it is instructive to see how fares an alternative approach based on a 5D metric like (1) above. That metric has a 4D part which is scaled by a length $L$ that was identified as the Compton wavelength $h / mc$ of a test particle with spacetime coordinates $x^\gamma$. The 4D part of the 5D metric defines the line element of spacetime as usual via $ds^2 = g_{\alpha\beta} \left( x^\gamma \right) dx^\alpha dx^\beta$. Then it is possible to define a dimensionless action, which can be used to obtain a wave function:

$$I \equiv \int ds / L = \int \left( h / mc \right)^{-1} ds \quad , \quad \psi = \exp\left( iI \right) \quad . \tag{2}$$

The first derivative of this gives

$$p_\alpha = \left( h / i\psi \right) \partial \psi / \partial x^\alpha \qquad , \qquad (3)$$

where the 4-momenta are defined as usual $\left( p^\alpha \equiv mc\,u^\alpha = mc\,dx^\alpha / ds \right)$. The second derivative, taken covariantly if the spacetime is curved, splits into a real part and an imaginary part. One of these gives $p^\beta_{;\beta} = 0$, the standard conservation law for the momenta. The other part gives

$$\Box\psi + \left( c / h \right)^2 m^2 \psi = 0 \qquad , \qquad (4)$$

where $\Box\psi \equiv g^{\alpha\beta}\psi_{,\alpha;\beta}$ (a comma denotes the partial derivative and a semicolon denotes the co-variant derivative). This last relation is the Klein-Gordon equation.

The same relation (4), it should be noted, can also be obtained by considering the full dynamics of the 5D metric as given by the components of the 5D geodesic equation [8]. Indeed, wavelike motion is a general property of canonical metrics like (1) if they have a timelike, as opposed to spacelike, extra dimension. Both choices are allowed since the extra dimension does not have the physical nature of a time. (In noncompactified 5D field theory with metrics that are not of canonical type, the extra coordinate is associated with a scalar field which is connected to particle mass.) The most general form of the canonical metric, with a shift $\left( l_0 \right)$ along the $l$ axis, has line element

$$dS^2 = \left[ \left( l - l_0 \right) / L \right]^2 ds^2 + \varepsilon dl^2 \qquad . \qquad (5)$$

Here $ds^2$ refers to the 4D metric of any vacuum solution of Einstein's equations, where $\varepsilon = +1$ for $\Lambda < 0$ (as above) and $\varepsilon = -1$ for $\Lambda > 0$. It is by now well known that 5D null paths given by $dS^2 = 0$ correspond to 4D paths with $ds^2 = 0$ for photons *and* $ds^2 > 0$ for massive particles [1, 8]. Using the null-path condition in (5) shows that the metric has two modes: an oscillatory one for $\varepsilon = +1$ and a monotonic one for $\varepsilon = -1$. With (*i*) to indicate the two cases, they are given by

$$l = l_0 + l_* \exp\left[ \pm (i) \, s / L \right] \quad . \tag{6}$$

Here $l_*$ is an arbitrary constant, which is the amplitude of the wave for the oscillatory mode. For the monotonic mode, $L = h / mc$ as before where *m* is the mass of the particle. The implications of metrics like (5) and paths like (6) are currently under investigation, particularly in regard to the properties of the vacuum [2, 8]. It is, of course, tempting to argue that the oscillatory and monotonic modes of the canonical metric be applied to the problem of wave-particle duality. However, this meets a problem which is of long standing in the foundations of quantum mechanics: Do the wave and the particle exist as separate things, or are they aspects of the *same* thing? The exact solution (1) describes only a wave, and as such has $g_{44} > 1$ and $\Lambda < 0$ in the canonical metric. Conversely, the corresponding particle has $g_{44} < 1$ and $\Lambda > 0$. These solutions are separate, and each implies significant spacetime curvature. The curvature contributions would only cancel if both modes were to occur simultaneously, as the result of an horizon in the fifth dimension or a change in the mode of the complex scalar field associated with it [1]. As far as the solution (1) is concerned, vacuum waves by themselves would only be expected to occur in certain high-energy environments, such as the early universe or localized regions today. The knotty problem of wave-particle duality in 5D is beyond the scope of the present account, but will be taken up in future work.

## 4. Discussion and Conclusion

There is a solution (1) of the field equations of 5D relativity which describes a wave in the classical vacuum. The latter has positive pressure but negative energy density, and accordingly the phase velocity of the wave may exceed the speed of light. It is difficult to classify these waves, but they are presumably not conventional gravitational waves, which exist in truly empty space ($\rho = p = 0$) and have velocities limited by the speed of light. The new waves exist in ordinary 3D space and have properties which more closely resemble those of the de Broglie waves of old quantum theory. 5D relativity is actually compatible with 4D wave mechanics, notably in regard to the Klein-Gordon equation (4). Other canonical-type 5D metrics, which embed vacuum solutions of the 4D Einstein equations, also have wave-like and particle-like modes. The implication is that the puzzling properties of de Broglie waves in 4D are due to truncating a well-behaved higher-dimensional solution.

De Broglie waves are sometimes viewed with suspicion, largely because they entail velocities that exceed light speed. However, waves of this type arise inevitably in any 5D theory of the kind used in unification, if the extra coordinate is timelike. Consider flat 5D space with line element $dS^2 = c^2 dt^2 - \left( dx^2 + dy^2 + dz^2 \right) + dl^2$. For the 5D null-path $dS^2 = 0$, the apparent velocity along the x-axis has magnitude $dx/dt = \left[ c^2 + \left( dl/dt \right)^2 \right]^{1/2}$, and can exceed $c$. For more realistic situations, in which there is a significant vacuum measured by a cosmological constant $\Lambda < 0$, superluminal waves are a natural consequence. From the viewpoint of 5D field theory, waves of de Broglie type have to be considered real.

This conclusion will, of course, not be surprising to those involved in modern experiments on the wave nature of matter [9]. Recent refinements of the classic double-slit experiment, and other setups, have confirmed the wave nature of matter for relatively large molecules (e.g., carbon-60). For a composite object, the de Broglie wavelength depends on its constituents, and there is a regular transition of the wave properties from massive particles to entangled photons. Yet despite extensive experimental work, the origin of these waves is not well understood from the theoretical side. What has been shown here is that waves of de Broglie type can be regarded as the natural manifestation of a higher dimension of the kind needed for the unification of the interactions of physics. If such vacuum waves were present in the early universe, or exist in certain environments today, they could provide a way to test for one or more extra dimensions.

Acknowledgements

This paper concentrates on certain issues discussed in the review at arXiv: 1205.4452. Thanks go to J.M. Overduin for checking the solution (1) of the text on the computer, and to other members of the Space-Time-Matter group (http://5Dstm.org).

gives the speed of the wave as observed in $F$ as $w \equiv x / t = c^2 / v$.  That is, $vw = c^2$ as stated in the main text.